          [1999/12/02 1.01 (PWD)]
\documentclass[a4paper,twocolumn]{esapub} 
\usepackage{natbib}
\usepackage{psfig}

\title{TEMPORAL EVOLUTION OF {\it f}-MODE FREQUENCIES AND RADIUS}
\author{Kiran Jain}
\author{S. C. Tripathy}
\author{A. Bhatnagar}
\affil{Udaipur Solar Observatory, Physical Research Laboratory, P.O. Box No. 198, 
Udaipur 314 004 (INDIA)}

\begin{document}


\maketitle

\begin{abstract}
We have analysed temporal evolution in centroid frequencies and splitting 
coefficients of solar {\it f}-modes  obtained from MDI/SOHO. The data were 
divided into 20  sets covering a period from May 1, 1996 to August
31, 2000.  The variation in frequencies is estimated to be 68 nHz over the period 
of four years which includes the rapidly rising phase of the solar cycle 23. 
This change is much smaller than that observed for {\it p}-mode frequencies. It is 
also noticed that the {\it f}-mode frequencies appear to be weakly correlated with 
solar activity indices as compared to the {\it p}-mode frequencies. We have also
inferred the relative change in the solar radii and notice a 1 year periodicity
which may be associated with the solar cycle variation. 
\end{abstract}

\section{Introduction}

With the availability of precise frequencies from both Global Oscillation Network Group (GONG)
 and  Solar Heliospheric Observatory (SOHO)
instruments, refined analysis of solar cycle dependent parameters are being
carried out. In these studies, {\it p}-mode frequencies have been widely used 
since these modes penetrate deep inside the sun and thus provide a way to infer changes beneath the 
surface. In contrast, {\it f}-modes, which are  the surface gravity modes,
are less well studied. Most of the efforts in {\it f}-mode studies have been concentrated in estimating the
seismic radius of the sun since these mode frequencies are independent of the 
stratification in the solar interior \citep{dzi98,dzi00,antia00}.
Frequency splittings obtained from {\it f}-modes have also been used to study the zonal flows which are
associated with torsional oscillations \citep{howard80} as seen in BBSO data \citep{wod93}.
 These sub-surface flows were first noticed by \citet{sasha97} in the Michelson Doppler Imager (MDI) 
 {\it f}-mode splittings. This analysis was later extended by \citet{schou99} for a longer time period
which showed average drift of the zonal flows towards the equator. \citet{toomre00} have confirmed 
these zonal flows using both {\it p}- and {\it f}-mode splittings derived from MDI 
data. 

Clearly, we are still lacking a basic understanding as to 
what causes the frequency changes. Keeping this in view, we analyse the {\it f}-mode 
frequencies and splitting coefficients obtained 
from the MDI medium-$\ell$ program for the period which includes the rising phase
of the solar cycle 23. We look for possible solar cycle changes in the mean frequencies
and even order splitting coefficients. These are further correlated with  activity
indices. We have also estimated the variations in the seismic radius
for the period covered in this study. 
		
\section{The Data}
The results presented here consist of twenty 72 day data sets obtained with the MDI 
instrument on board SOHO and 
covers a period from May 1, 1996 to August 31, 2000 with two breakdowns in between
(see Schou, 1999 for more details). Each data set consists of 
centroid frequencies $\nu_{n,l}$ and 36 splitting 
coefficients $a_k$. For the present work, only {\it f}-modes
are used which samples a region very close to the surface. In particular,  the 
analysis is restricted to 41 common modes in the degree range of 
217 $\le \ell \le$ 286 and frequency range of 
 1484 $\mu$Hz $\le \nu \le$ 1702 $\mu$Hz unless otherwise mentioned. 

\section{Analysis and Results}
\subsection{Variation of the Sun's Eigen Frequencies}
\begin{figure}
\centering
\begin{center}
\epsfxsize=2.8in \epsfbox{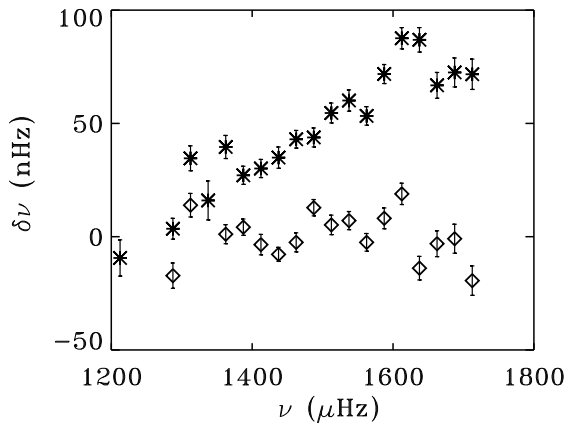}
\vspace{0.5cm}
\caption{The binned frequency shifts for two  independent periods: set 1 (May 1 -- July 7, 1996) represented by 
diamonds and set 20 (June 21 -- August 31, 2000) by stars.  The 
shift has been calculated with reference to set 6 (April 26 -- July 6, 1997). The error bars represent
mean error in shift. \label{fig1}}
\end{center}
\end{figure}

The variation of $\ell$ averaged frequency shift as a function of frequency 
for low (May 1 -- July 7, 1996) and high (June 21 -- August 31, 2000) activity
period with reference to set 6 (April 26 -- July 6, 1997) is shown in Figure~\ref{fig1}. 
During the minimum activity period, $\delta\nu$ seems 
to be independent of $\nu$ while during high activity period, the frequency shift 
 increases sharply with frequencies.
The mean shift $\delta\nu$ for a given  $\ell$ is calculated from the relation 
\begin{equation}
\delta\nu(t)  = \frac{ {\sum_{\ell}\frac{\delta\nu_{\ell}(t)}{\sigma_{\ell}^2}}}{{\sum_{\ell}\frac{1}{\sigma_{\ell}^2}}}
\end{equation}  

\input epsf 

\begin{figure}
\centering
\begin{center}
\input epsf
\epsfxsize=2.8in\epsfbox{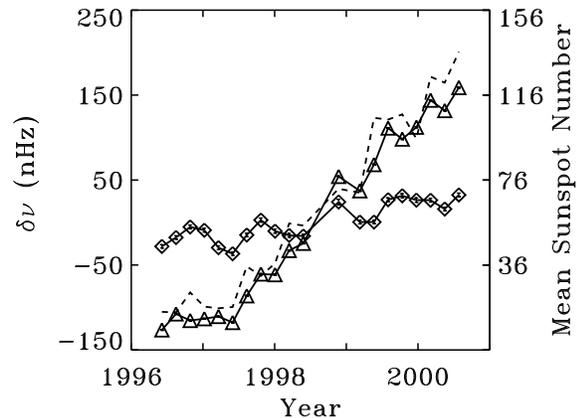}
\vspace{0.5cm}
\caption{The temporal evolution of frequency shifts for {\it p}- ( line joined by triangles)
and {\it f}-modes (line connected by diamonds). The shift has been calculated with 
respect to mean of all 20 data sets separately. The dashed line  represents the 
solar activity levels denoted here by the International sunspot number. For clarity,
the error bars are not shown.  
\label{fig2}}
\end{center}
\end{figure}

where $\sigma_{\ell}$ is the error in the frequency measurement. 
The  variation of $\delta\nu$ with time  is shown 
in Figure~\ref{fig2}. Here we notice an oscillatory behaviour with a periodicity
of about one year. 
A similar result is also found by \citet{antia2000b}. 
The temporal evolution of {\it p}-mode frequencies (1500 $\mu$Hz $\le \nu \le$ 3500
$\mu$Hz) and the activity index
represented by the International sunspot number $R_I$ is also shown 
in the same figure. It is
evident that the change in {\it p}-mode frequencies is 
larger than that of {\it f}-mode frequencies. It is further clear that the {\it p}-mode
frequency shifts follow the activity index quite closely while the mean frequency 
shifts of {\it f}-modes appear to be weakly correlated (also see Table~1).

To study the correlation between mean frequency shifts and solar activity cycle, we have considered 
the following different activity indices:
$R_I,$ the International 
sunspot number obtained from the Solar Geophysical Data (SGD); KPMI, Kitt Peak 
Magnetic Index  which represents the line of sight magnetic 
field setrength and averaged over the full disk from Kitt Peak magnetograms; 
 MPSI, Magnetic Plage Strength Index from Mount Wilson 
magnetograms which represents the absolute values of the magnetic field
strengths between 10 and 100 gauss; MWSI, Mount Wilson Sunspot Index which represents the 
magnetic field strengths above 100 gauss;  He~I, equivalent width of He~I 
10830 \AA ~line, averaged over the whole disk from Kitt peak and   
$F_{10}$, integrated radio flux at 10.7 cm from 
SGD. 
Both  Pearson's ($r_p$) and 
Spearman's ($r_s$) correlation coefficients are given in Table~1.

It is observed that in case of {\it f}-modes, the activity indices have 
 weak correlation with frequency shifts. 
  In comparison,  {\it p}-modes display higher correlation coefficients showing stronger correlation 
with activity indices as was earlier pointed out by \citet{ab00} and \citet{hkh99}. However, it is 
interesting to note that MWSI with other magnetic indices
 has strong correlation with {\it f}-modes while it has the 
weakest correlation with {\it p}-modes. 
  Thus, it appears that MWSI which represents the average magnetic field 
strenth above 100 gauss behaves differently with 
  {\it p}- and {\it f}-modes
and needs to be investigated more closely as the solar cycle progresses. 
The observation that {\it f}-modes in comparison with {\it p}-modes
are weakly correlated with magnetic field indices  
further confirms  that magnetic fields are present at deeper layers rather than near
 the solar surface. 
\begin{table}
  \begin{center}
    \caption{Correlation statistics for {\it f}- and {\it p}-mode frequency shifts.}\vspace{1em}
    \renewcommand{\arraystretch}{1.2}
    \begin{tabular}[h]{lcccccc}
      \hline
      Activity & \multicolumn{3}{c}{{\it f}-mode}&\multicolumn{3}{c}{{\it p}-mode}\\
      \cline{2-3} \cline{5-7}
      Index& $r_p$     &  $r_s$     &&& $r_p$& $r_s$      \\
      \hline
      $R_I$ & 0.84 & 0.80&&& 0.98&0.93 \\
      KPMI  & 0.88& 0.78 &&& 0.99 & 0.94  \\
      MPSI & 0.87& 0.77&&&0.99 &0.95  \\
      MWSI  & 0.87&0.82&&&0.91&0.83  \\
      F$_{10}$& 0.87&0.78&&&0.99&0.96  \\
      He I& 0.85 & 0.79&&& 0.99&0.97\\
      \hline \\
      \end{tabular}
    \label{tab:table1}
  \end{center}
\end{table}

\subsection{Seismic Radius}
Since the early nineteen century, the temporal variation in the Sun's diameter 
has been reported by many observers \citep{lac96,noeel97}.
 The recent measurements report
a change of 0.1 arcsec to 1 arcsec in the measured semi-diameter which implies 
a change of 70 to 700 km in the radius. Assuming that the change in 
frequencies is entirely due to the change in radius and there is no contribution 
from the magnetic field, convection etc., 
the average 
frequency difference is related to change in radius by 
\begin{equation}
<\frac{\delta\nu}{\nu}> = - \ \frac{3}{2} \ \frac{\delta R}{R} ,
\end{equation}
where $\delta\nu$ is the centroid frequencies calculated earlier and the differences 
are taken with respect to the mean of all 20 data sets. The  evolution of the relative differences 
in the 
seismic radius is shown in Figure~3. The maximum change is 
found to be 29.8 km over a period of four years, the seismic radius decreasing 
with increase of activity lavels and thus indicating anti-correlation between them. 
A relative variation in the solar radius of the order of 6$\times$10$^{-6}$ was first 
reported by \citet{dzi98} which corresponds to approximately 4 km over 
a period of 2 years around the solar minimum. If this variation is correlated with solar
activity, we would expect much larger change, as reported in this paper,  
between minimum and maxium activity
period. However, the analysis of \citet{dzi00} did not find 
any systematic trend of correlation with sunspot number. 
 A similar study using GONG data \citep{antia00} for the period
1995 to 1998 reported a decrease in solar radius by about 5 km, during the period in
which the solar activity was  increasing. 
It has been known that
the high wave number modes are known to deviate significantly
from the simple dispersion relation \citep{duv98,ab99} and the inclusion 
of higher degree 
in our analysis  may cause this large change in the solar radius. 
\input epsf

\begin{figure}
\centering
\begin{center}
\epsfxsize=2.8in\epsfbox{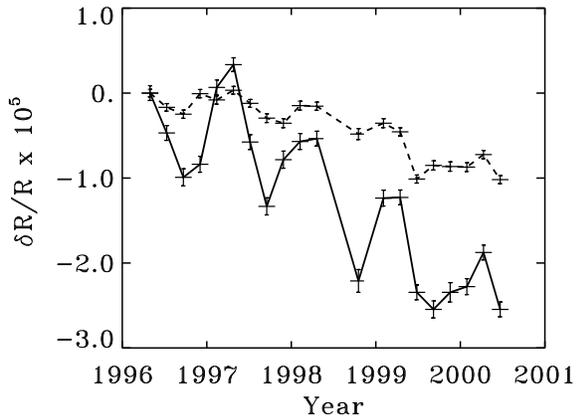}
\vspace{0.5cm}
\caption{Relative differences in the radius of the sun as inferred from
{\it f}-mode frequencies. The solid line is the result as derived from frequency tables
containing splitting coefficients up to $a_{36}$ and covers the degree range of 
217 to 286. The dashed line shows the result 
using the frequency tables containing splitting coefficients 
up to $a_6$ and includes modes of 100 $\le$ $\ell$ $\le$ 200 only. \label{fig3}}
\end{center}
\end{figure}

In order to check the stability of our results, we have repeated the analysis
by considering the frequency tables which contains the splitting coefficients 
only up to $a_6$ and thus lowering the $\ell$ values to 119. The relative 
difference in radius in this case is  found to be 14 km and clearly demonstrates
the degree dependence. We have also estimated the relative difference in seismic
radius from activity minimum to maximum by restricting the common modes between
 100 $\le$  $\ell$ $\le$ 200 and this is  shown as dashed line in Figure~3. In this case, the 
 change in radius is found to be  7 km, consistent with the results of
 \citet{antia00} which used the same range of $\ell$ values. 
 \begin{figure}
\centering
\begin{center}
\input epsf
\epsfxsize=2.8in\epsfbox{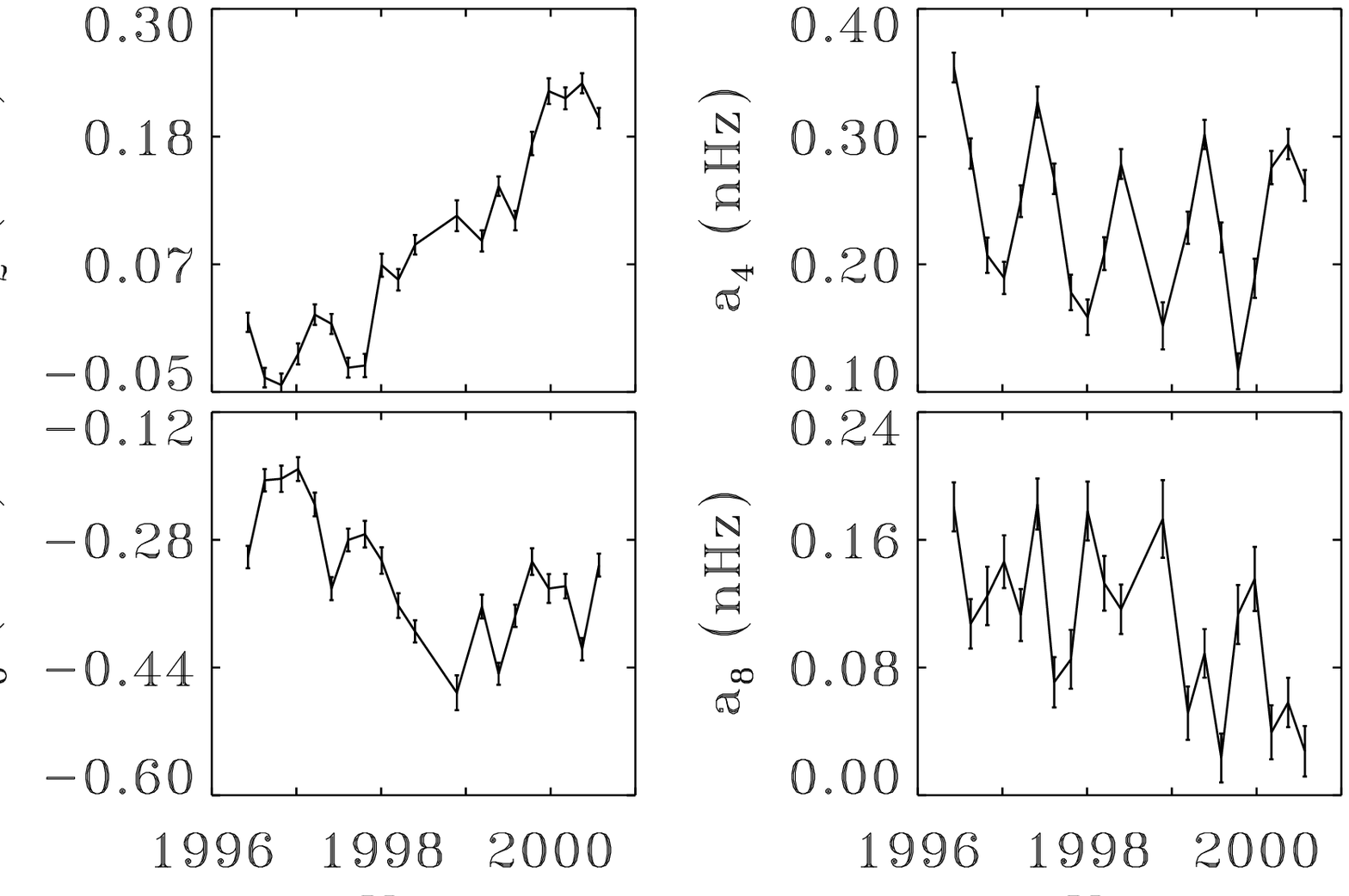}
\epsfxsize=2.8in\epsfbox{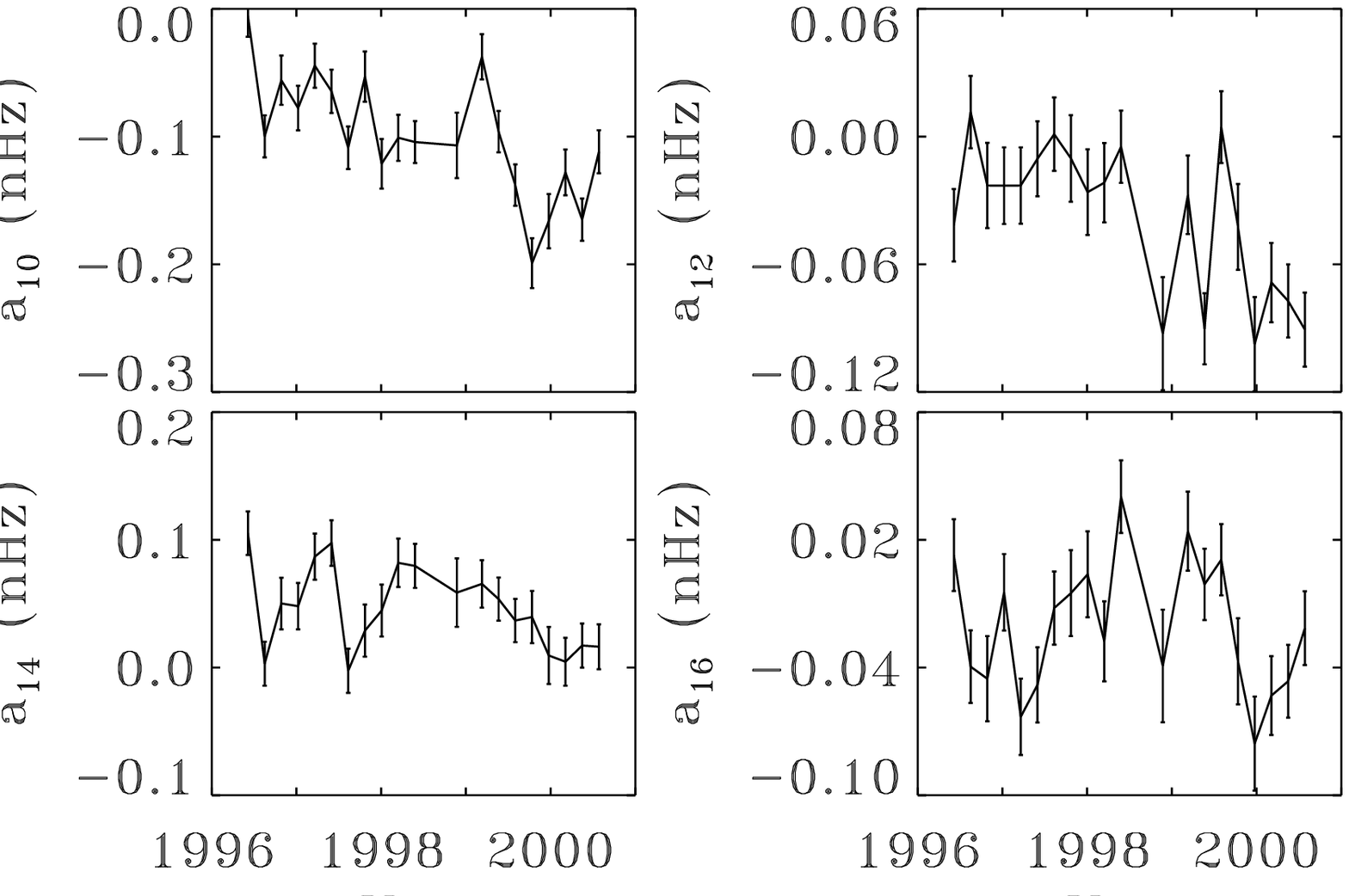}
\vspace{0.5cm}
\caption{Average even order splitting coefficients as a function of index and time. For 
each index $i$, the $a$-coefficients are joined with lines denoting temporal
evolution. Error bars are 1 $\sigma$. The averaging was done only over common modes. 
\label{fig4}}
\end{center}
\end{figure}
 
 Since the change in seismic
radius  reflects the relative differences in frequencies, the oscillatory pattern visible 
in Figure~2 persists in Figure~3. However, it is not clear if this period is associated with 
solar cycle variations since the data used here spans only 4 years. Also, this period is 
very close to the orbital period of the earth and one may expect systematic errors of this 
period being introduced in the data. 

\subsection{Even Order Splitting Coefficients}
 The variation of even order coefficients up to $a_{16}$ for {\it f}-modes are shown in Figure~4.  
These coefficients seem to be uncorrelated with solar cycle except for $a_2$.  
In contrast, the even order splitting coefficients of  {\it p}-modes are known to be correlated 
with solar cycle. In particular, \citet{hkh99} find significant correlation between the even $a$-coefficients
and the corresponding Legendre coefficients of a fit to the Kitt Peak magnetograms. A similar 
result based on the BBSO Ca K plage data for higher order was recently reported by \citet{dzi00}. 
 
\section{conclusions}  
We have analysed  the {\it f}-mode frequencies and splitting coefficients obtained from 
MDI medium-$\ell$ program for the period May 1996 to August 2000. 
We find that the mean shift in these mode frequencies are weakly correlated with 
activity indices as compared to the {\it p}-mode frequencies.
In case of MWSI, which represents the magnetic field above 100 Gauss, the {\it f}-modes 
have the best correlation while in case of {\it p}-modes, the correlation is weakest. 
We have also 
estimated the relative difference in the seismic radius as due to the change in frequencies 
and find that 
the change is a function of degree of the mode. If we consider only those modes with 
degrees between 100 and 200,
we find the change in solar radius from minimum to maximum of the solar cycle is approximately 
7 km and is one order smaller than that is expected from the analysis of MDI data by 
\citet{dzi98}. If we use  the high degree modes $\ell$ up to 286 which is 
approximately the same as used by \citeauthor{dzi98}, we find a significantly higher value. 
We also report here a 1 year periodicity 
in both the {\it f}-mode frequencies and seismic radius, alhough it is not clear if these are 
associated with  the solar cycles or reflects the systematic errors in the data.   

\section*{Acknowledgments}
 
This work utilises data from the Solar Oscillations Investigation / Michelson Doppler Imager
 on the Solar and Heliospheric Observatory. We would like to thank J. Schou for providing us 
the MDI data. 
 SOHO is a mission of international cooperation
 between ESA and NASA. NSO/Kitt Peak magnetic, 
and Helium measurements used 
here are produced cooperatively by NSF/NOAO; NASA/GSFC and NOAA/SEL.  This work is 
partially supported under the CSIR Emeritus Scientist Scheme and Indo--US collaborative 
programme--NSF Grant INT--9710279.

\end{document}